\documentstyle[psfig,a4wide,cite]{article} 

\psfigurepath{figs}

\begin{document}

\title{Entropy-based analysis of the number partitioning problem}
\author{A.R. Lima$^{1,2}$ and M. Argollo de Menezes$^2$\cite{arlima}\\ 
  $^1$ Laboratoire de Physique et M\'echanique des Milieux
  H\'et\'erog\`enes, ESPCI Paris \\ 10 rue Vauquelin, 75231 Paris
  Cedex 05, France \\ $^2$ Instituto de F\'{\i}sica, Universidade
  Federal Fluminense \\ Av. Litor\^anea 24210-340, Niter\'oi, RJ,
  Brazil}
\date{Version: 12/07/2000 - Printed: \today}
\maketitle

\begin{abstract}
  In this paper we apply the multicanonical method of statistical
  physics on the number-partitioning problem (NPP). This problem
  is a basic NP-hard problem from computer science, and can be
  formulated as a spin-glass problem. We compute the spectral
  degeneracy, which gives us information about the number of solutions
  for a given cost $E$ and cardinality $m$.  We also study an
  extension of this problem for $Q$ partitions. We show that a
  fundamental difference on the spectral degeneracy of the generalized
  ($Q>2$) NPP exists, which could explain why it is so difficult to
  find good solutions for this case. The information
  obtained with the multicanonical method can be very useful on the
  construction of new algorithms.
\end{abstract}



\section{ Introduction }
\label{sec:intro}

The use of statistical mechanics tools to understand the main ideas
underlying problems as diverse as biological, social and economic
systems has become a common task, both theoretically and
computationally \cite{peliti91,oliveira99}. Recently, these tools have
been applied to computer science problems
\cite{cheeseman91,gent95,mertens98,ferreira98,monasson99,ferreira99,ferreira99a,mertens00},
not intending to solve them exactly, but to understand their
complexity and the underlying mechanisms generating such complex
behavior. In this paper we focus on the number partitioning problem, a
fundamental problem in theoretical computer science ~\cite{garey79}.
Our aim is to apply the multicanonical method (MUCA) ~\cite{berg91}
of statistical physics on this problem and study the behavior of
nearly optimal solutions. This information is
important for the development of new algorithms which try to find
optimal solutions.

In the next section we discuss the number partitioning problem and its
formulation as a spin glass problem. We also introduce the
multi-partitioning problem and map it onto a $Q$-states Potts model.  In
section \ref{sec-esm} we present the Lee formulation of the
multicanonical method and apply it to our problem. On section
\ref{sec-conclusions} we discuss our results both for the classical
and for the multi-partitioning problem.

\section{The number partitioning problem}
\label{np-problem}

The number partitioning problem (NPP) is, according to Garey and
Johnson \cite{garey79}, one of the six basic computer science
problems. Given a set $A=\{a_1, a_2, a_3, a_4,..., a_N\}$ with $N$
integer numbers, the traditional NPP consists of partitioning the set
$A$ into two disjoint sets $A_1$ and $A_2$ such that the difference
\begin{equation}
  E = \left|\sum_{a_i \in A_1} a_i - \sum_{a_i \in A_2} a_i\right|
\label{eq-unpp}
\end{equation}
is minimized. If there are $N_1$ numbers in the set $A_1$ and $N_2$
numbers in the set $A_2$, then
\begin{equation}
m=\left| N_1-N_2 \right |
\end{equation}
is called the \emph{cardinality} of the set.

On the unbalanced NPP the only condition is to minimize the cost
function (eq.~\ref{eq-unpp}) without any restriction to the value of
$m$. The problem of finding good solutions (whenever they exist) for
the unbalanced (non-fixed $m$) NPP was essentially solved by the
deterministic Karmakar-Karp-Korf complete algorithm
\cite{karmarkar82,korf98}. This algorithm was generalized by Mertens
for the balanced ($m$ fixed) case \cite{mertens99}. Some recent papers
addressed the possibility of carrying a statistical analysis of this
problem ~\cite{mertens98,ferreira98,ferreira99,ferreira99a,mertens00}
obtaining interesting results, such as: the existence of an
easy-to-hard transition \cite{mertens98}, the non self averaging
property of the ground state energy \cite{ferreira98}, the analytical
derivation of the lower bounds for the energy as a function of the
cardinality ~\cite{ferreira99} and the equivalence of the NPP to a
random cost problem~\cite{mertens00}.

For that purpose, it was proposed a mapping of the NPP problem onto a
spin-glass model: associate to each number $a_i$ a new variable $s_i$
(which we call ``spin'') such that if $a_i \in A_1$ them $s_i=-1$,
otherwise $s_i=+1$. With this mapping, we can search for a configuration of
spins $s_1,...s_N$ which minimizes the cost function (or energy)
\begin{equation}
\label{eq-spinglass}
E = \left|\sum_{i=1}^{N}s_{i}a_{i}\right|
\end{equation}
or its square,
\begin{equation}
E_{SG}=\sum_{ij}^{N}J_{ij}s_{i}s_{j}
\end{equation}
with $J_{ij}=a_ia_j$, which we recognize as an infinite-range
spin-glass Hamiltonian. We can also write down the cardinality on a
``magnetization-like'' fashion
\begin{equation}
m = \frac{1}{N} \left|\sum_{i=1}^{N}s_{i}\right|.
\end{equation}
Finding an optimal solution for the number partitioning problem
consists of finding the spin configuration of the fundamental state on
the spin-glass problem. This is a very difficult task, mainly because
of the great number of metastable states separated by a hierarchy of
increasingly high energy barriers \cite{mezard87}.

A much more difficult problem is multi-partitioning.  This problem
consists of partitioning the set of numbers $A$ into $Q$ disjoint sets
such that the energy and magnetization associated to these partition
assume the values $E$ and $m$, respectively. This problem has several
applications \cite{coffman91,tsai92}, such as the division of $N$ different
jobs (computer programs) into $Q$ computers.

We can map this problem onto a Potts spin-glass by assigning to each
number $a_i$ a spin $s_i$ which can assume integer values from $1$ to $Q$.
These spin magnitudes represent the set to which the number belongs.
Hence, the energy can be written as
\begin{equation}
E = \sum_{i=1}^{Q}\sum_{j>i}^{Q}\left|\epsilon_{i}-\epsilon_{j}\right|
\end{equation}
where $\epsilon_i = \sum_{k=1}^{N}a_k\delta_{(s_{k},i)}$ is the sum of the
elements in the set $i$. In the same way we define the magnetization as
\begin{equation}
m = \frac{1}{N}\frac{\sum_{i=1}^{Q}\sum_{j>i}^{Q}\left|n_{i}-n_{j}\right|}{Q-1}
\end{equation}
where $n_i = \sum_{k=1}^{N}\delta_{(s_{k},i)}$ is the number of elements in
the set $i$.

Clearly this problem is much more complex than the traditional NPP
($Q=2$). 

In the next section we show that the
multicanonical method can be used to determine the spectral
degeneracy of the problem, i.e., the number of solutions $g(E,m)$
which have a given energy $E$ and magnetization $m$. In the
statistical mechanics sense, this completely characterizes the problem
since the (adimensional) entropy is given by $S(E,m)=\ln g(E,m)$.

\section{The multicanonical method (Entropic Sampling)}
\label{sec-esm}
The multicanonical method was introduced in 1991 by Berg and
Neuhaus\cite{berg91}, and the basic idea of this method is to sample
micro-configurations of a given system performing a biased random walk
(RW) in the configuration space which leads to another unbiased random
walk (i.e. with uniform distribution) along the energy axis. This walk
must have a visiting probability of each energy level $E$ which is
inversely proportional to $g(E)$, the quoted spectral degeneracy. If
one can measure the transitions probabilities from an energy level $E$
to all other energy levels, one is able to obtain $g(E)$. The
multicanonical method has been shown to be very efficient in obtaining
satisfactory results for $g(E)$ in a large variety of problems such as
evolutionary problems \cite{choi97}, phase equilibrium in binary lipid
bilayer \cite{besold99} and optimization problems~\cite{lee94} (for
reviews of the method see \cite{mucareviews}).  The Entropic Sampling
Method (ESM) \cite{lee93}, which we will use throughout this paper,
has been proven to be an equivalent formulation of MUCA \cite{berg95}.
Here we are interested in the multiparametric formulation of the
multicanonical method, since we must obtain the spectral degeneracy
$g(E,m)$ \cite{shteto97} as a function of two parameters $E$ and $m$.
Let $E(X)$ and $m(X)$ be the energy and magnetization associated to
the microstate $X$, the transition probability between two states
$X_i$ and $X_f$ is given by
\begin{equation}
\label{estransition}
\tau(X_i, X_f) = e^{-[S(E_f,m_f)-S(E_i,m_i)]} 
               = \frac{g(E_i, m_i)}{g(E_f,m_f)}
\end{equation}
where $S(E,m)=\ln g(E,m)$ is the entropy, $E_i=E(X_i)$($E_f=E(X_f)$)
is the energy of the initial (final) state, $m_i=m(X_i)$($m_f=m(X_f)$)
is the magnetization of the initial (final) state and $g(E,m)$ is the
spectral degeneracy. The transitional probability (eq.
\ref{estransition}) satisfies a detailed balance equation and leads to
a distribution of probabilities where a state is sampled with
probability $\propto 1/g(E,m)$. The successive visitations along the
energy axis follow a uniform distribution, but unfortunately $g(E,m)$
is not known {\em a priori}. One way of obtaining $g(E,m)$ is to
construct it iteratively, such as in the ``Entropic Sampling'' method
proposed by Lee ~\cite{lee93}. This method can be summarized as
follows:

{\em Step 1:} Start with $S(E,m)=0$ for all states;

{\em Step 2:} Perform a few\footnote{when compared with the number of
  steps which will be done in the step 4} unbiased RW steps in the
configuration space and store $H(E,m)$, the number of tossed movements
to each state with energy $E$ and magnetization $m$ (in this stage all
movements are accepted);

{\em Step 3:} Update $S(E,m)$ according to

\begin{equation}
S(E,m) = \left\{ \begin{array}{ll}
         S(E,m) + \ln H(E,m) & \mbox{, if $H(E,m) \neq 0$} \\
         S(E,m)              & \mbox{, otherwise.}         
        \end{array}
        \right. 
\label{esentropy}
\end{equation}

{\em Step 4:} Perform a much longer MC run using the transitional
probability given by eq. (\ref{estransition}), storing $H(E,m)$.

{\em Step 5:} Repeat 3 and 4. This is considered one iteration.

\begin{figure}[thpb]
  \centerline{\psfig{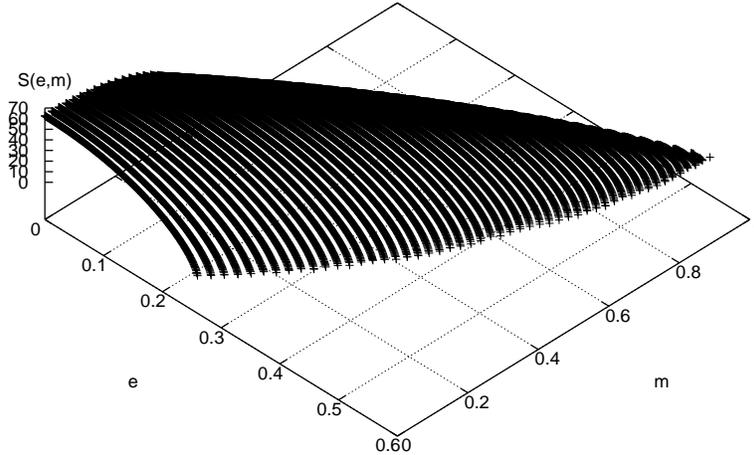}}
\caption{Typical entropy curve for the case of bi-partitioning.
  Here the entropy in computed for a single instance of 100 numbers
  chosen randomly between $0$ and $10^{10}$. The energy is normalized
  by the largest possible value $E_{max}=10^{12}$.}
\label{fig-entropy}
\end{figure}

Usually, the number of Monte Carlo steps used in step 4 increases with
the number of iterations. For a detailed description of the method see
refs. \cite{lee93,shteto97,newman99}.  We applied this algorithm to
the problem of bi- and multi-partitioning in order to obtain the
entropy $S(E,m)$, which is shown for the case of bi-partitioning in
figure \ref{fig-entropy}.  These results were obtained for a single
instance (disorder realisation) of $100$ integer numbers chosen
randomly between $0$ and $10^{10}$. All results shown are typical ones.
It is important to stress that, after obtaining the entropy of the
system through extensive simulations, all thermodynamic averages for
different cardinalities $m$ can be executed with no need to any
further computer effort.

In the next section we are going to analyze this entropy in order to
recover well-known results concerning bounds of the $(E,m)$ curve for
nearly optimal solutions. Through analysis of the entropy of the
multi-partitioning problem it will be clear that the change of
complexity of this problem for $Q>2$ is associated to fundamental
changes of the entropy curve.

\section{Numerical results}
\label{sec-conclusions}

\begin{figure}[thpb]
  \centerline{\psfig{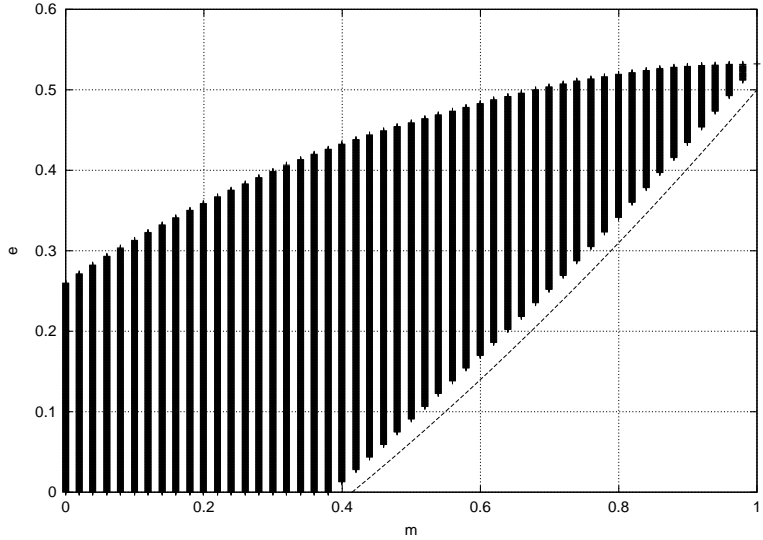}}
\caption{Collapse of all entropies $S(e,m)$ on the $(e,m)$ plane,
  evidencing the upper and lower bounds for $e$ as a function of the
  cardinality $m$. Again, $e=E/E_{\rm max}$. The dashed line is the
  analytical prediction given by Ferreira and
  Fontanari~\cite{ferreira99}.  Typical results for a single instance
  with $N=100$ numbers.}
\label{fig-bounds}
\end{figure}

In ~\cite{ferreira99}, Ferreira and Fontanari have
calculated through the replica trick analytical estimates for the
average lower bound of the energy $E$ as a function of magnetization
(cardinality) $m$. By collapsing the $z-$axis on the $(x,y)$ plane
(here the $(E,m)$ plane) we are able to see this limit and also the
upper bound for $E$. In figure \ref{fig-bounds} we show our numerical results
and the analytical prediction of Ferreira and Fontanari. The energy is
normalized appropriately.

\begin{figure}[thpb]
  \centerline{\psfig{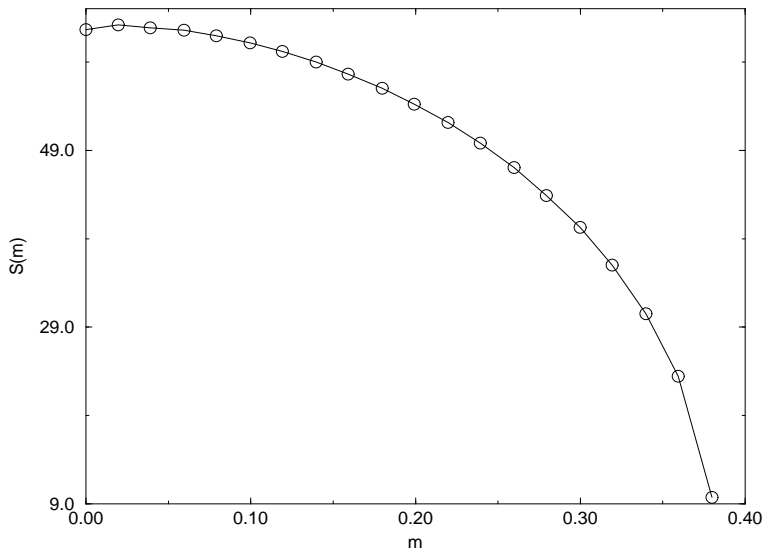}}
\caption{Entropy of an $N=100$ instance as a function of the
  cardinality for nearly optimal solutions. The maximum number of
  solutions does not correspond to an equipartition of the set, but to
  two subsets with $N/2-1$ and $N/2 +1$ numbers, respectively. This
  result is not characteristic of a single instance (disorder
  realisation), but seems to appear on most of the numerical results.
}
\label{fig-em}
\end{figure}

In computer science one is only interested in the optimal solutions,
that is, the information contained on the first ``slice'' of the
$S(E,m)$ surface, the $S(0,m)$ plane. In figure \ref{fig-em} we show
$S(\epsilon,m)$, where $\epsilon$ is our numerical tolerance, which we
chose to be $(E_{MAX}-E_{MIN})/1024$.

One interesting feature we have observed numerically is that the
maximum number of solutions does not occur for $m=0$; it is
easier to find solutions where the number of elements on each set is
not exactly equal. This feature is not characteristic of a particular
instance.

\begin{figure}[thpb]
  \centerline{\psfig{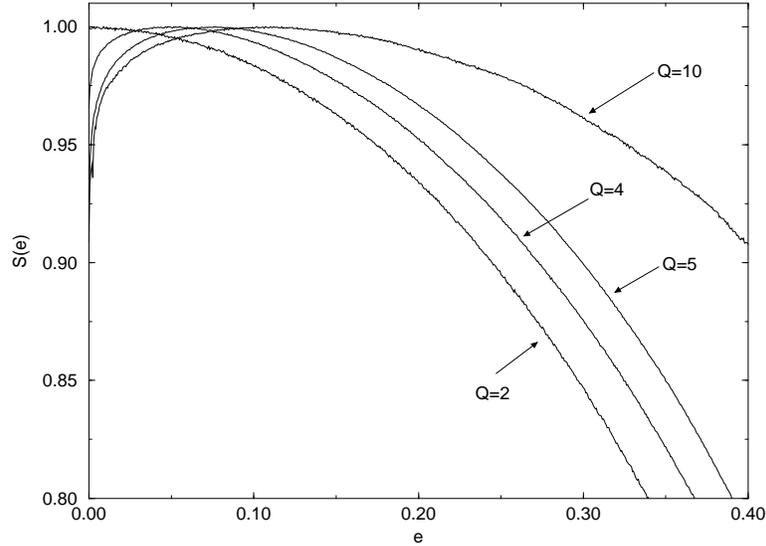}}
\caption{The normalized entropy of the system as a function of the
  energy $e$ for different number of partitions $Q$. The fundamental
  difference between bi- and multi-partitioning is that for the latter
  the maximum of the entropy deviates from the origin, which turns out
  to be a minimum, making it more difficult to find nearly optimal
  solutions.}
\label{fig-q}
\end{figure}

Now we show our results for the multi-partitioning problem, where no
reasonable deterministic algorithm that finds optimal solutions exists
for any instance. As far as we know, there is no theoretical study of
the NPP for $Q>2$. If we look at the number of solutions $g(E)=\sum_m
g(E,m)$ (or the entropy $S(E)=\ln g(E)$) for different cardinalities,
we observe a fundamental difference between $Q=2$ and $Q>2$ results
(figure \ref{fig-q}).  For the case $Q=2$, we confirm recent results
obtained by Mertens \cite{mertens00}, who showed that the traditional
number-partitioning problem ($Q=2$) is essentially equivalent to the
random cost problem, the problem of finding the minimum in an
exponentially long list of random numbers, where no other algorithm
than random search can be more efficient. As the maximum of the
entropy lies near $E=0$ in the NPP ($Q=2$) problem (fig.
\ref{fig-entropy}), any algorithm based on random movements will drive
the system close to the fundamental state.

For $Q>2$ we have a completely different scenario. The maximum of the
entropy is not near $E=0$, indeed, $E=0$ is a minimum and the number
of solutions with $E\approx 0$ decreases, at least, exponentially with
$Q$. This means that any algorithm based on random movements would
drive the system away from the fundamental state on the $Q>2$ case.
The effects of this behavior for the construction of new algorithms
have to be taken into account. On the traditional differencing scheme
\cite{karmarkar82,korf98,mertens99} we can say that, if the number of
nearly optimal solutions increases exponentially as $E\rightarrow 0$,
it is always possible to find better and better solutions, the
computational time spent on the search being the only barrier. For the
$Q>2$ NPP this kind of procedure does not seem to work, since the
number of solutions \emph{decreases} exponentially for $E\rightarrow
0$.

\section{Conclusions}
\label{sec:conclusions}

We showed in this paper that the multicanonical method for obtaining
thermodynamic averages of statistical systems can provide a tool for
assessing the complexity of computer science problems, such as the
number-partitioning problem (NPP). This problem is one of the six
basic computer science problems, according to Garey and Johnson
\cite{garey79}, and can be formulated as a spin-glass problem. Based
on this analogy we proposed a statistical mechanics method for
computing the spectral degeneracy of the NPP problem which gives us
information about the number of solutions for a given cost $E$ and
cardinality $m$. We have also studied an extension of this problem for
$Q$ partitions, where there exists no good deterministic algorithm
which finds optimal solutions. We observed a fundamental difference
between the classical ($Q=2$) and the generalized ($Q>2$) NPP, which
explains why it is so difficult to find good solutions for the latter
case. This information can be very useful in the construction of new
algorithms.

\section*{Acknowledgments}
The authors wish to thank the International Center for Theoretical
Physics (ICTP) in Trieste/Italy for financial support.  This work has
begun while attending the ``School on Statistical Physics Methods
Applied to Computer Science'' at ICTP. We also thank F.~F.~Ferreira
and J.~F.~Fontanari for useful discussions and J.F.~Stilck for a
careful reading of the manuscript.  Both authors acknowledge financial
support from Brazilian agencies CNPq and CAPES.

\end{document}